\documentclass[fleqn]{aa}  

\usepackage{graphicx}
\usepackage{xcolor}
\usepackage{hyperref}
\usepackage{txfonts}
\usepackage{calligra}
\usepackage{calrsfs}
\usepackage[T1]{fontenc}
\usepackage[mathcal]{eucal}
\usepackage{longtable}
\newcommand{\vectX}{\bf {\it X}}
\newcommand{\vectTheta}{\bf {\it \Theta}}

\newcommand{\matrGamma}{\bf \Gamma}

\begin{document}

   \title{On the correlation between dark matter, intracluster light and globular cluster distribution in SMACS0723.}

   \titlerunning{WSLAPping SMACS0723}
  \authorrunning{Diego et al.}

   \author{J.M. Diego
         \inst{1}\fnmsep\thanks{jdiego@ifca.unican.es}
       \and
        M. Pascale\inst{2}
       \and
        B. Frye\inst{3}
       \and
        A. Zitrin\inst{4}
       \and
       T. Broadhurst\inst{5,6,7}
       \and
       G. Mahler\inst{8,9}
       \and 
       G.B. Caminha\inst{10}
       \and
       M. Jauzac\inst{8,9,11,12}
       \and 
       Myung Gyoon Lee\inst{13} 
       \and 
       Jang Ho Bae\inst{13} 
       \and 
       In Sung Jang\inst{14} 
       \and 
       Mireia Montes\inst{15,16}
    }      
   \institute{Instituto de F\'isica de Cantabria (CSIC-UC). Avda. Los Castros s/n. 39005 Santander, Spain
        \and
         Department of Astronomy, University of California, 501 Campbell Hall \#3411, Berkeley, CA 94720, USA
         \and
         Department of Astronomy/Steward Observatory, University of Arizona, 933 N Cherry Ave., Tucson, AZ 85721, USA
          \and
          Physics Department, Ben-Gurion University of the Negev, P. O. Box 653, Be’er-Sheva, 8410501, Israel
          \and
         Department of Physics, University of the Basque Country UPV/EHU, E-48080 Bilbao, Spain
         \and
          DIPC, Basque Country UPV/EHU, E-48080 San Sebastian, Spain
          \and
          Ikerbasque, Basque Foundation for Science, E-48011 Bilbao, Spain
          \and
          Centre for Extragalactic Astronomy, Durham University, South Road, Durham DH1 3LE, UK
          \and
          Institute for Computational Cosmology, Durham University, South Road, Durham DH1 3LE, UK
          \and
        Max-Planck-Institut für Astrophysik, Karl-Schwarzschild-Str. 1, D-85748 Garching, Germany
          \and
          Astrophysics Research Centre, University of KwaZulu-Natal, Westville Campus, Durban 4041, South Africa
          \and
          School of Mathematics, Statistics \& Computer Science, University of KwaZulu-Natal, Westville Campus, Durban 4041, South Africa 
          \and
          Astronomy Program, Department of Physics and Astronomy, SNUARC, Seoul National University, 1 Gwanak-ro, Gwanak-gu, Seoul 08826, Republic of Korea
          \and
          Department of Astronomy \& Astrophysics, University of Chicago, 5640 South Ellis Avenue, Chicago, IL 60637, USA
          \and
         Instituto de Astrofísica de Canarias, c/ Vía Láctea s/n, E-38205 La Laguna, Tenerife, Spain
         \and
         Departamento de Astrofísica, Universidad de La Laguna, E-38205 La Laguna, Tenerife, 
          }

 \abstract{
    We present a free-form model of SMACS0723, the first cluster observed with JWST. This model makes no strong assumptions about the distribution of mass (mostly dark matter) in the cluster and we use it to study the possible correlation between dark matter with the intracluster light and distribution of globular clusters. To explore the uncertainty in mass modelling, we derive three lens models based on spectroscopically confirmed systems and new candidate systems with redshifts predicted by the lens model derived from the spectroscipic systems. 
    We find that beyond the radius of influence of the BCG, the total mass does not trace the ICL, implying the need for a dark component (dark matter). Two loop-like structures observed in the intracluster light do not have an obvious correspondence with the total mass (mostly dark matter) distribution. The radial profiles of the ICL and the distribution of globular clusters are similar to each other, but steeper than the profile of the lens model. More specifically, we find that the total mass is shallower by 1 dex in log scale than both ICL and globular cluster profiles. This is in excellent agreement with N-body simulations of cold dark matter.  
    
   }
   \keywords{gravitational lensing -- dark matter -- cosmology
               }

   \maketitle
%

\section{Introduction}
 After its launch on December 25th 2021, on July 11st 2022, the first color image from James Webb Space Telescope (JWST) was presented to the world. The image showed a view of the distant infrared universe with a detail and depth never seen before at these wavelengths. This image was centered on a massive galaxy cluster at $z=0.39$, SMACS J0723.3-7327 (or SMACS0723 hereafter) acting as a powerful gravitational lens. This natural lens magnifies the galaxies in the background. Some of these background galaxies appear repeated several times in the image, since they take different paths which are later refocused by the gravitational lens into the JWST telescope. 
 
In anticipation for these first JWST data, a lens model based on HST data and listing the first few sets of multiple images and some spectroscopic redshifts for them was posted on the arXiv by \cite{Golubchik2022} on the same date (July 11th). The JWST data became itself public on July 13th 2022, and just a day after the data release, two papers presenting new candidates to multiply lensed galaxies and new lens models were submitted simultaneously to arXiv \citep{Mahler2022,Pascale2022}.\footnote{The difference between the two submission times was just 13 seconds!}. A day after these two papers appeared on arXiv, a third one was submitted presenting an additional lens model and new lensed system candidates \citep{Caminha2022}. Other papers focusing on the high-redshift galaxies lensed by SMACS0723 and their properties quickly followed \citep{Ferreira2022,Cheng2022,Laporte2022,Adams2023,Carnall2023}.
This frenzy over the new data reflects the excitement and anticipation of the community for the new JWST data. JWST is revolutionizing the field of astronomy in a similar fashion as it was done by its predecessor, the Hubble Space Telescope (HST), at the end of the 20th and beginning of the 21st centuries.

 The first image of JWST reveals approximately two dozen lensed system candidates, five of which have spectroscopic redshift estimations from MUSE and JWST data \citep{Golubchik2022,Sharon2022,Pascale2022,Mahler2022,Caminha2022}\footnote{As this paper was being finished, a new spectroscopic redshift for system 4 (z=2.211) is provided in \cite{Noirot2022}}. One of the surprises in the new data is the presence of hundreds of point like sources near the large member galaxies in the cluster (possibly stripped galactic nuclei or compact globular clusters) \citep{Lee2022,Faisst2022}.
Some of the lensed galaxies also show small unresolved structures which could be compact star forming regions, globular clusters, groups of stars or even individual stars in cases of extreme magnification \citep{Mowla2022}. These can prove very valuable in upcoming works, which will look for flux anomalies between pairs of counterimages \citep{Pooley2012,Chan2020}. The unresolved nature of these substructures, together with the large magnification of some of them can be used to study models of dark matter (DM) which predict anomalous flux ratios between these pairs of images.  
An additional surprise in the new data is the unusual distribution of the intracluster light (or ICL hereafter), already noted in \cite{Pascale2022} and \cite{Mahler2022} and studied in more detail in \cite{Montes2022}. 
The ICL is formed by stars not bound to any galaxy of the cluster, but to the gravitational potential of the cluster as a whole \citep[see][for a review]{Montes2022b}. 
ICL is observed to be older near the centre of the clusters \citep{Montes2018}, suggesting an earlier accretion of the central ICL region. 
The ICL is particularly interesting in the context of DM. Similarly to DM particles, the stars responsible for the ICL (as well as the globular clusters and galactic core remnants) can be considered as non-interacting particles that respond only to gravity. Hence, one would expect a tight correlation between the distribution of ICL and DM \citep{Montes2019,Asensio2020}. In the case of SMACS0723, the ICL departs from the expected smooth distribution predicted for DM from N-body simulations \citep{Asensio2020} and shows two loop-like structures at $\approx 200$ kpc from the central BCG, one to the east and one to the west of the cluster core. These loop-like structures may be the result of a relatively recent merger but given the expected connection between the stars in the ICL and the DM it is interesting to study if similar structures can be found in the distribution of dark matter from the lens model. \\

The previous lens models rely on some parameterization of the mass distribution, usually by placing ellipsoids at the positions of galaxies, and/or large elliptical halos near the center of the cluster to account for the contribution from DM, or by assuming it follows the cluster galaxy distribution. Hence, they are less than ideal to study the possible correlation between the DM and ICL distributions. In this paper, we present an additional lens model based on a free-form technique that makes no assumptions about the underlying distribution of DM. A comparison between the DM and ICL (or globular cluster) distributions can then be done without being subject to assumptions made about the distribution of DM.

\begin{figure*} 
   \includegraphics[width=18cm]{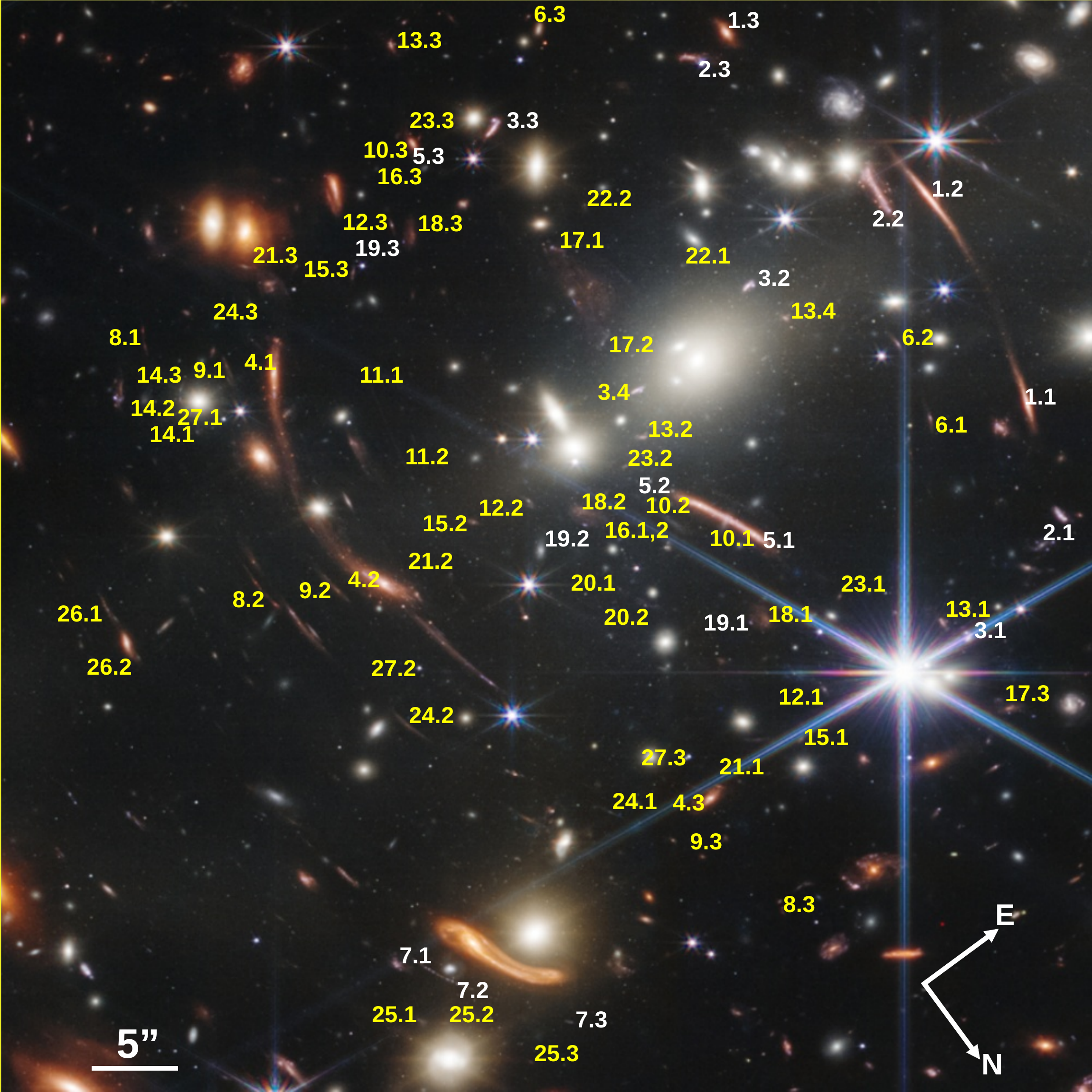}

      \caption{Central $\approx 1$ arcminute region of SMACS0723 with systems of lensed galaxies. Systems in white have spectroscopic redshifts and are the ones used to build the driver model or Model-1. Systems in yellow do not have spectroscopic redshifts but are used in combination with  the spectroscopic systems to build the lens Model-2 and Model-3. Note the hundreds of unresolved sources surrounding the BCG. These are mostly globular clusters and galactic core remnants. Unless otehrwise noted, all figures in this plot are in the same orientation as this one. 
         }
         \label{Fig_SMACS0723}
\end{figure*}

The paper is organized as follows.
In section \ref{Sect_constraints}, we discuss the lensing constraints used to derive the lens model. 
Section \ref{Sect_WSLAPplus} gives a brief introduction to the free-form algorithm used to derive the lens model, making no assumptions about the distribution of DM. 
In section \ref{Sect_lensmodel}, we present the driver model, or Model-1,  which is derived using only lensed systems with known redshifts (spectroscopic).
Section \ref{Sect_SLz} uses the driver model to make predictions for the redshifts of the candidate lensed systems without spectroscopic redshifts. These redshifts are an interesting alternative to (and some times are more precise than) the more common photometric redshifts.  
In section \ref{Sect_Lensmodels2}, we use the lens model predicted redshifts and present two additional models (Model-2 and Model-3), which use all the additional systems with constrained redshifts and, in the case of Model-3, increases also the spatial resolution of the DM component. Model-2 and Model-3 are useful to explore the uncertainty in the lens model due to i) the lens system definition (spectroscopic sample vs full sample), and ii) the spatial resolution in the lens model.  
In section~\ref{Sect_ICL}, we study the correlation between the ICL, globular cluster distribution and DM. We discuss our results and present our conclusions in section~\ref{Sect_discussion}. We adopt a standard flat cosmological model with $\Omega_m=0.3$ and $h=0.7$. At the redshift of the lens (z=0.39), and for this cosmology, one arcsecond corresponds to 5.29 kpc.

\section{Lensing constraints}\label{Sect_constraints}
The lensing constraints used in this work are compiled from the three most recent works discussed previously \citep{Pascale2022,Mahler2022,Caminha2022}. Positions, and IDs of these systems are presented in Table A.1 in appendix A. When possible, we maintain the original ID of earlier work. Candidates 6.3 and 16.3 in \cite{Pascale2022} are updated with the nearby candidates 7.3 and 11.3 respectively from \cite{Mahler2022}. For convenience, Table A.1 also includes the IDs used in earlier works. Among the systems in this table, five of them have spectroscopic redshifts. All systems are marked in Figure~\ref{Fig_SMACS0723} with circles and their corresponding ID. The five systems with spectroscopic redshifts are highlighted in bold.  \\

In addition to the classic lensing constraints, we add the position of critical curves which can be determined from the radial arc in system 5 (at z=1.425), and the merging pair of images in system 7 (at z=5.17). The positions of these critical points are added at the end of table A.1 and labeled CP5 and CP7 respectively. Each critical point contributes with two constraints as detailed in \cite{Diego2022_Godzilla}. Since at a critical point the magnification diverges, this can be easily incorporated by applying a rotation to the data by the angle determined by the elongation of the arc. After this rotation one can simply impose that the inverse of the tangential magnification equals zero, or similarly $1=\kappa-\gamma$. The second constraint is simply $\gamma_2=0$, which is satisfied when the rotation is applied.     

\section{WSLAP+}\label{Sect_WSLAPplus}
To optimize the lens model we use the code WSLAP+ \citep{Diego2005,Diego2007,Sendra2014,Diego2016}. A lens model derived using WSLAP+ is considered a hybrid type of model as it combines a free-form decomposition of the lens plane for the smooth large-scale component with a small-scale contribution from the member galaxies. Details can be found in references above. Here we give a brief description of the method. \\

We start with the classic definition of the lens equation
\begin{equation} 
\beta = \theta - \alpha(\theta,\Sigma) \, , 
\label{eq_lens} 
\end{equation} 
where $\theta$ is the observed position of the source, $\alpha$ is the deflection angle, $\Sigma(\theta)$ is the unknown surface mass-density of the cluster at the position $\theta$, and $\beta$ is the unknown position of the background source. 
The optimization of the WSLAP+ solution takes advantage of the fact that the lens equation can be expressed as a linear function of the surface mass density, $\Sigma$. WSLAP+ parameterizes  $\Sigma$ as a linear superposition of functions, which translates into $\alpha(\theta,\Sigma)$ being also linear in $\Sigma$. 

In WSLAP+, the surface mass density, $\Sigma$, is described by the combination of two components; 
i) a smooth component (usually parameterized as superposition of Gaussians) corresponding to the free-form part of the model, or large scale cluster potential; and 
ii) a compact component that accounts for the mass associated with the individual galaxies in the cluster.  \\
For the smooth component we use Gaussian functions defined over a grid of points. A Gaussian function is simple and enables fast computation of the deflection field, but also provides a good compromise between the desired compactness and smoothness  of the basis function. 
The grid configuration can be defined as regular (all grid points have the same size) or irregular (grid points near the centre are in general smaller). Adopting a regular grid is similar to a flat prior in the mass distribution while an irregular grid can be interpreted as a model with a prior on the mass distribution with higher mass density assigned to smaller cells. 
Since one of the goals of this paper is to study the possible correlation between the DM distribution and the ICL, we adopt a regular grid, since this makes minimal assumptions about the mass distribution.\\

For the compact component, we directly adopt the light distribution in the JWST band F277W  around the brightest member elliptical galaxies in the cluster. 
For each galaxy, we assign a mass proportional to its surface brightness. This is the only free parameter. This mass is later re-adjusted as part of the optimization process. The number of parameters connected with the compact component depends on the number of adopted layers. Each layer contains a number of member galaxies. The minimum number of layers is 1, corresponding to the case where all galaxies are placed in the same layer, that is, they are all assumed to have the same mass-to-light ratio. In this case, the single layer is proportional to the light distribution of all member galaxies, and is assigned a fiducial mass for the entire mass of the member galaxies. For each layer there is one extra parameter which accounts for the renormalization constant multiplying the map of the mass distribution, that is optimized by WSLAP+. For the particular case of SMACS0723, we use 4 layers. The first layer contains the main BCG, the second layer contains a large elliptical galaxy $\approx 9"$ west of the main BCG. The third layer contains two large elliptical galaxies near the Beret galaxy discussed in \cite{Mahler2022}. Finally, the fourth layer contains all remaining member galaxies.  Member galaxies are selected from the standard red-sequence, and we also make sure that spectroscopic members identified in \cite{Mahler2022} are included in this set.  \\


As shown by \cite{Diego2005,Diego2007}, the strong and weak lensing problem can be expressed as a system of linear equations that can be represented in a compact form, 
\begin{equation}
\vectTheta = \matrGamma \vectX, 
\label{eq_lens_system} 
\end{equation} 
where the measured strong lensing observables (and weak lensing if available) are contained in the array $\vectTheta$ of dimension $N_{\Theta }=2N_{\rm sl}$ (plus $2N_{\rm wl}$ if weak lensing data is available), the unknown surface mass density and source positions are in the array $\vectX$ of dimension 
\begin{equation}
N_{\rm X}=N_{\rm c} + N_{\rm l} + 2N_{\rm s}, 
\label{eq_Nx}
\end{equation}
and the matrix $\matrGamma$ is known (for a given grid configuration and fiducial galaxy deflection field) and has dimension $N_{\Theta }\times N_{\rm X}$.  $N_{\rm sl}$ is the number of strong lensing observables (each one contributing with two constraints, $x$, and $y$), $N_{\rm c}$ is the number of grid points (or cells) that we use to divide the field of view, $N_l$ is the number of layers ($N_l=4$ in our case as mentioned above), and $N_s$ is the number of background sources being strongly lensed (each source represent two unknowns in $X$, $\beta_x$, and $\beta_y$). 

The solution, $X$, of the system of equations \ref{eq_lens_system} is found after minimizing a quadratic function of $X$ \citep[derived from the system of equations \ref{eq_lens_system} as described in ][]{Diego2005}. The minimization of the quadratic function is done with the constraint that the solution, $\vectX$, has to be positive. Since the vector $\vectX$ contains the grid masses, the renormalization factors for the galaxy deflection field and the background source positions, and all these quantities are always positive (the zero of the source positions is defined in the bottom left corner of the field of view).  Imposing  $\vectX>0$ helps constrain the space of meaningful solutions, and to regularise the solution, as it avoids unwanted large negative and positive contiguous fluctuations. 
A detailed discussion of the quadratic algorithm can be found in \cite{Diego2005}. For a discussion of its convergence and performance (based on simulated data), see \cite{Sendra2014}.

As discussed in \cite{Diego2022_Godzilla}, critical points can also be added as extra constraints. We identify two such constraints in systems 5 (at z=1.425) and system 7 (at z=5.1727) with spectroscopic redshifts, and include them in our set of lensing constraints. The addition of these two points act as anchors for the lens model, enforcing the critical curve to pass through the desired point at the given redshift.

\section{Driver lens model}\label{Sect_lensmodel}
Using the constraints listed in Table A.1, we first derive the driver model, or Model-1. This model is only based on systems with spectroscopic redshifts. For the case of SMACS0723, and at the time of writing these paper, five systems are known to have spectroscopic redshifts\footnote{As noted earlier, a new spectroscopic redshift was recently made available for system 4 in \cite{Noirot2022} at the time of finishing this paper. This new redshift (z=2.211) was not used in our analysis where we adopted our geometric redshift estimate (z=2). The difference in redshift is small and is not expected to have any significant impact in our results.}. These are marked in bold in Table A.1. For the grid, we use a regular distribution of $20\times20=400$ grid points. Given the relatively small number of lensing constraints, a significantly larger number of grid points results in nonphysical solutions with large mass fluctuations. 


Based on the 5 lensed systems with spectroscopic redshift, the driver model can be used to predict the redshift of the other system candidates in Table A.1. We do so in the next section.

\section{Redshifts predicted by the lens model}\label{Sect_SLz}

Using the driver model, we derive redshifts for all systems listed in table \ref{tab_Arcs}. The probability of a system to be at redshift $z$ is computed by;
\begin{equation}
    P(z) = exp(-V(z)/(2\sigma^2)),
    \label{eq_Pz}
\end{equation}
where $V(z)$ is the variance between the arc positions of a given system projected on the source plane at redshift $z$. The projection is done with the deflection field of the driver model (computed at redshift $z=3$) which is re-scaled to the desired redshift. The dispersion, $\sigma$, in the expression above is fixed to three pixels, or $\approx 0.18"$. This is a reasonable choice for well constrained systems, resulting in relatively narrow distributions for the redshift, and with uncertainty in the error prediction consistent with the observed error \citep[see for instance][for a more in depth analysis of the errors expected with this technique and for WSLAp+]{Diego2023}. 
Systems that are well reproduced by the driver model result in a small variance $V(z)$ near the optimal redshift, which in turn result in maximum values of $P(z)$ close to 1. Systems that are poorly reproduced by the driver model have larger values of $V(z)$, which reduce the maximum value of $P(z)$.
A low maximum probability for $P(z)$ does not necessarily mean that the system is a bad candidate. This can simply be the result of the driver model not being well constrained in that part of the lens system.  
Systems at high redshift tend to have broader probabilities, since for source redshift $z>2$ the deflection field varies slowly with redshift.

\begin{figure} 
   \includegraphics[width=9cm]{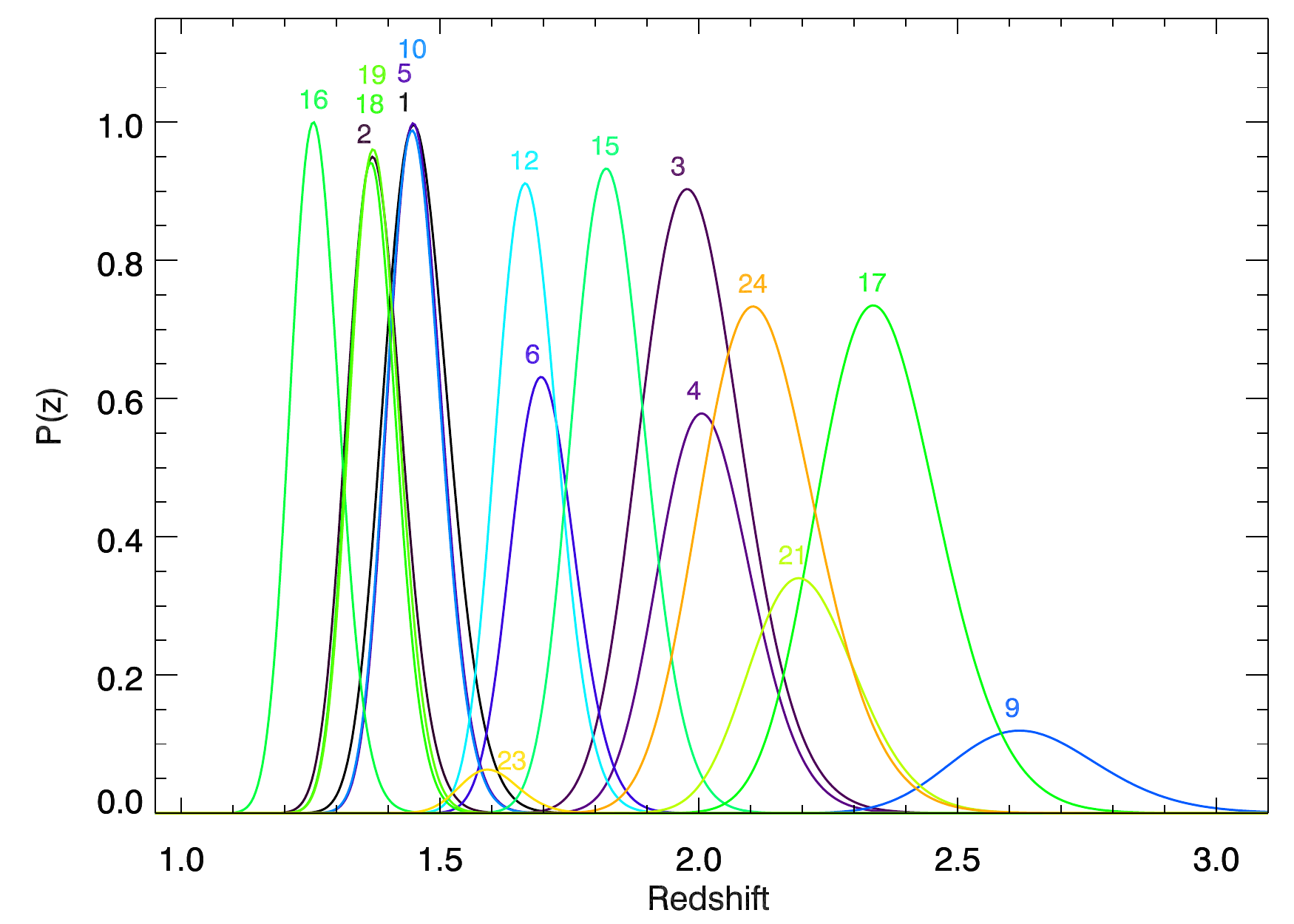}
      \caption{Redshifts predicted by the driver model for the case of well constrained systems. 
         }
         \label{Fig_SLzI}
\end{figure}

\begin{figure} 
   \includegraphics[width=9cm]{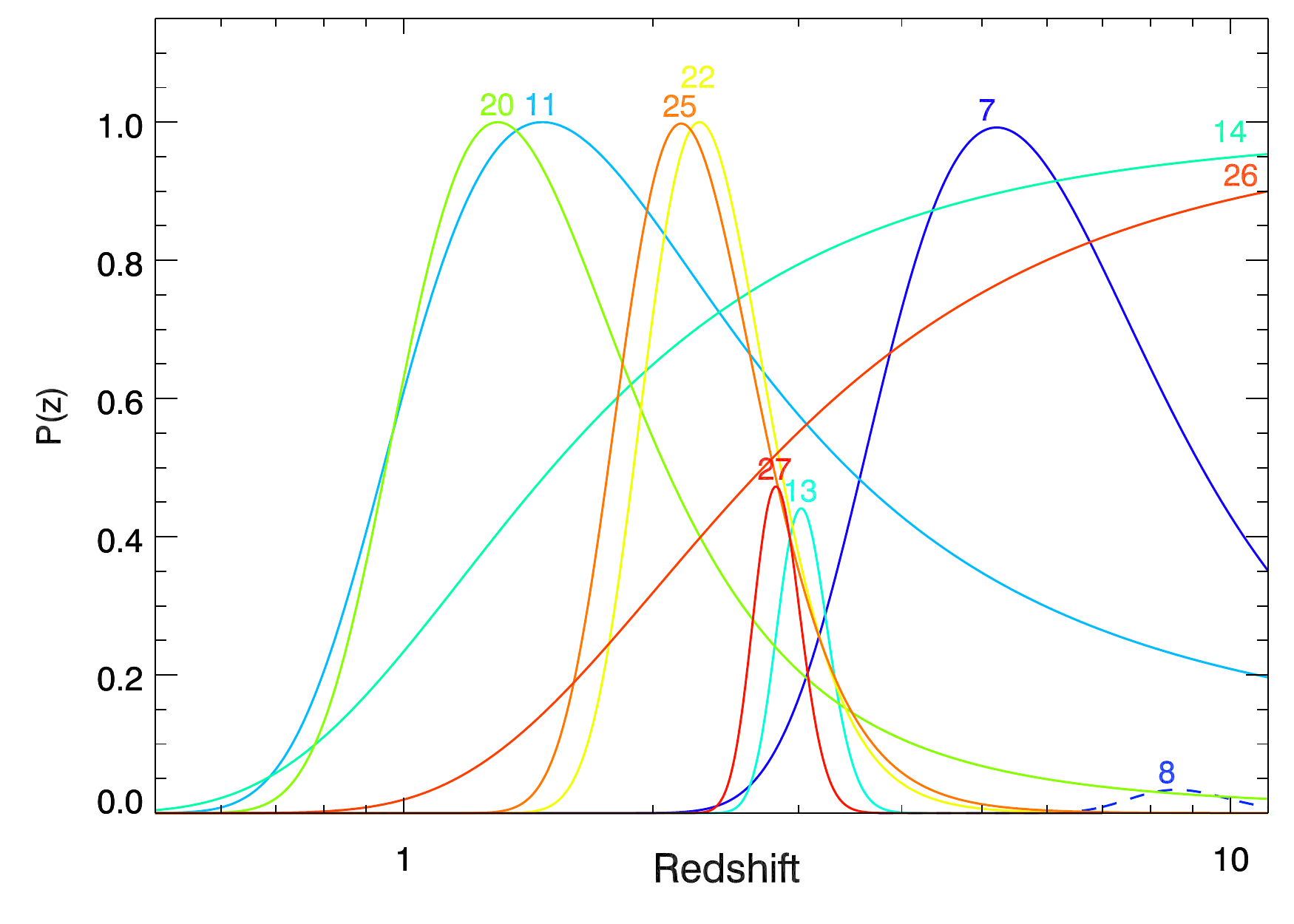}
      \caption{Redshifts predicted by the driver model for the case of poorly constrained systems. The probability of system  8, shown as a dashed line at $z\approx 8$, has been multiplied by a factor 100.   
         }
         \label{Fig_SLzII}
\end{figure}

The derived probabilities $P(z)$ can be divided in two groups. In the first group we find systems with well defined and relatively narrow probabilities. The probabilities for these systems are shown in Figure~\ref{Fig_SLzI}. Among these we find the systems with spectroscopic redshifts that were used to derive the driver model. Naturally, the maximum of $P(z)$ for these systems falls very close to the spectroscopic value. System 4 had its spectroscopic redshift estimated recently in \cite{Noirot2022} where they find $z=2.211$. As shown in Figure~\ref{Fig_SLzI}, the $P(z)$ for this system  contains the correct redshift within the 95\% confidence interval. \\
In a different group we find systems for which the redshift is not so well constrained. The probabilities for these systems is shown in Figure~\ref{Fig_SLzII}. Two systems (14 and 26) have no constrain on their redshift (z>13). The bad performance of these systems can be easily understood since they correspond to cases of galaxy-galaxy lensing, where the member galaxy acting as a lens is not optimized individually (these galaxies are part of layer 4 discussed in Section~\ref{Sect_WSLAPplus}). 

System 8 has a very low probability of $P(z)$. This probability is shown as a dashed line in Figure~\ref{Fig_SLzII}, and the probability has been multiplied by a factor 100, to make it visible in the figure. The low probability of system 8 can be interpreted as being a bad system or being in a region with poor constraints in the driver model. This is the case on the western part of the cluster, where system 8 lies, since only one system has a spectroscopic redshift in this region of the lens. 

Redshifts predicted by gravitational lenses are an interesting alternative to photometric redshifts, specially for high-redshift candidates for which photometry may be poor or nonexistent in more than one band (such as in the case of dropouts). A similar technique has been used in the past in other lenses with positive results and was recently applied in \cite{Diego2023} to predict the redshift of the new systems identified in the JWST data of El Gordo cluster.

\section{Full-sample lens models}\label{Sect_Lensmodels2}
\begin{figure} 
   \includegraphics[width=9cm]{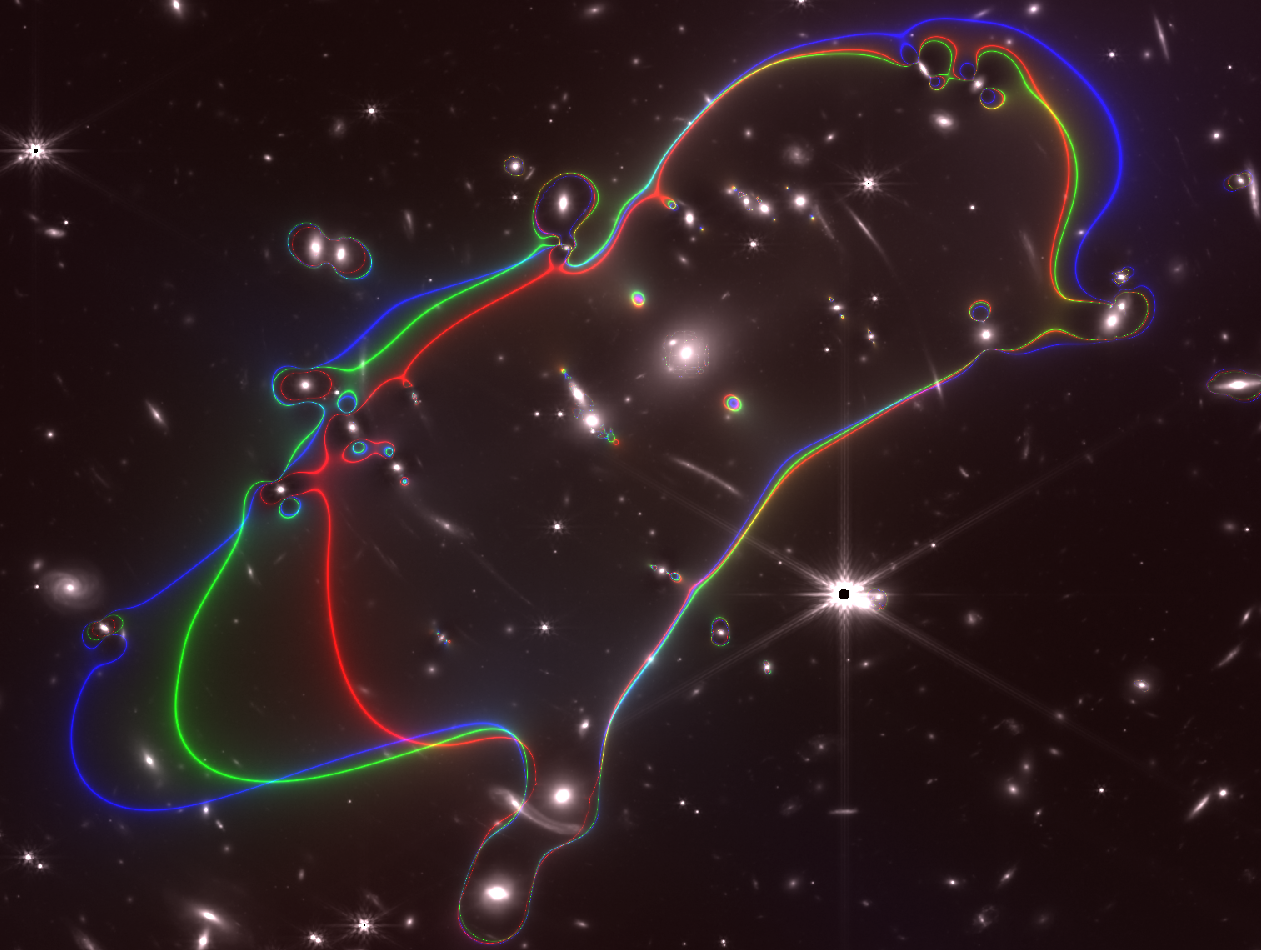}
      \caption{Critical curves of alternative lens models. All critical curves are computed at the redshift of system 7 (z=5.1727). The red curve corresponds to the driver model derived with the five spectroscopic systems and a grid of 20x20 points. The green curve uses the same grid configuration but is derived from the 25 systems with constrained redshifts. The blue curve uses the same 25 systems but is based on a higher resolution grid of 25x25 points. 
         }
         \label{Fig_AlternativeModels}
\end{figure}

Taking advantage of the redshifts predicted by the lens model discussed in the previous section,  we expand the number of constraints and update the lens model. As discussed above, the redshift for systems 14 and 26 cannot be constrained by the lens model so we exclude these two systems from our list of constraints. The remaining number of systems totals 25, and the number of constraints exceeds 150 (x and y positions of each arc plus the two critical point positions, each contributing also with two constraints). Using these constraints we derive two models. One that we refer to as Model-2, which is derived with the full set of constraints (excluding systems 14 and 26), and a regular grid of $20\times20=400$ points.  We increase the number of grid points to $25\times25=625$ in a third model that we refer to as Model-3. Increasing the number of grid points even further can result in unstable solutions. For instance, with a grid of $30\times30=900$ points we obtain a solution that places too much mass in the edges of the field of view and introduces relatively large mass fluctuations across the entire field so we do not consider solutions with more than 625 grid points. 
The critical curves for Model-2 and Model-3 are shown in Figure~\ref{Fig_AlternativeModels} as green and blue curves respectively. For convenience we include again in this figure the critical curve for the driver model (or Model-1) in red. All three curves are again computed at the redshift of system 7 ($z=5.1727$). The three models produce consistent results in the eastern part of the lens, which is the portion of the cluster where the number density of spectroscopic redshifts is the highest. In contrast, the critical curves in the west differ significantly from one another, indicating that the western part of the cluster is more poorly constrained. The addition of new spectroscopic systems in this part of the lens will reduce the uncertainty in the lens model. \\

In terms of mass, we can compare with previous published results based on parametric models. Both \cite{Mahler2022} and \cite{Caminha2022} quote the total projected mass within a cylinder of radius 128 kpc centered in the BCG. This radius corresponds approximately to the Einstein radius for a source at $z>2$, and it is the radius within which the lens model can be properly constrained with strong lensing data. They find masses of $8.26 \pm0.04 \times 10^{13} {\rm M}_{\odot}$ and  $8.7 \pm0.2 \times 10^{13} {\rm M}_{\odot}$ respectively, and within the aforementioned 128 kpc radius. For our three lens models we find  $7.28\times 10^{13} {\rm M}_{\odot}$, $7.31\times 10^{13} {\rm M}_{\odot}$, and $7.15\times 10^{13} {\rm M}_{\odot}$ for Model-1, Model-2 and Model-3 respectively, and within the same radius. This is approximately 10\% less than in the parametric models. 

\section{Correlation between the dark matter, intracluster light, and globular cluster distributions}\label{Sect_ICL}

\begin{figure} 
   \includegraphics[width=9cm]{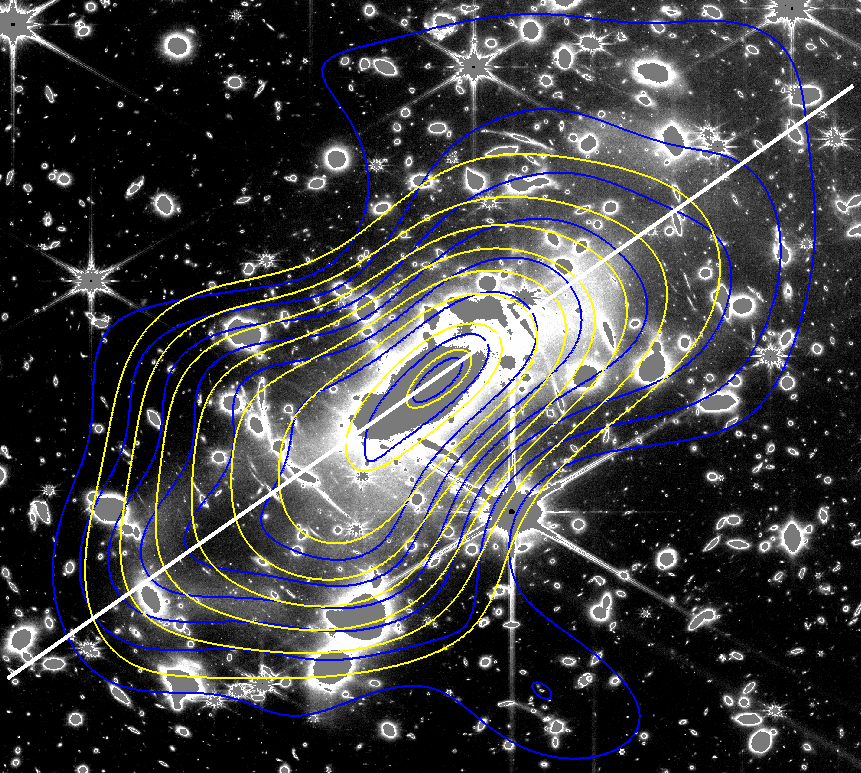}
      \caption{Projected total mass vs ICL. The contours represent the smooth component of the lens model obtained with the 25 constrained systems. The yellow contour is obtained with a regular grid of 20x20 cells while the blue contour is obtained with a higher resolution grid of 25x25 cells. The image is a masked version of the F277W band, where the ICL light can be better appreciated. The contours correspond to values of the convergence, $\kappa$, computed at a fiducial source redshift of $z_s=3$. Space between contours correspond to $\delta\kappa=0.1$, with values starting at $\kappa=0.5$. The last contour is for $\kappa=1.15$.  The white straight line marks the direction over which we construct the one-dimensional scan of the light profile and DM models.
         }
         \label{Fig_ICL}
\end{figure}

In all three models discussed in the previous sections we find that the critical curves for the three models are consistent among them, with the largest differences concentrating in the west portion of the cluster. Hence the lens model is relatively well constrained for different choices of lensed systems and grid configurations. In this section, we pay special attention to the distribution of light in the ICL, and the distribution of globular clusters. We are interested in the possible correlation between the ICL, globular cluster and the DM distributions. Figure~\ref{Fig_ICL} shows how the ICL presents two loop-like structures to the east and west of the cluster. At $\approx 200$ kpc from the central BCG and towards west, a cavity-like structure can be appreciated in the ICL \citep[Giant West Loop in ][]{Montes2022}. Although not as clear, a similar cavity can be also observed towards the east at approximately the same distance from the central BCG. This is a surprising feature in the ICL where one expects to find more uniform distributions. Recent merger activity can result in trails of stars being stripped away from their host galaxy by tidal forces.  These tails are however much smaller than the observed loops in the ICL of SMACS0723, and connect with the host galaxy. In the case of the loops of SMACS0723, one cannot establish any correspondence between the loops and a member galaxy. On the other hand, as noted by \cite{Mahler2022}, the offset between the radial velocity of the central BCG and the mean redshift of the cluster suggests a recent past merger (a relaxed cluster would have no offset), offering a possible explanation for the odd distribution of the ICL. \\

Whatever the cause for the morphology of the ICL, it is interesting to compare its two-dimensional distribution with the distribution of mass from our lens models. Since approximately 85\% of the projected mass of the cluster is expected to be dark matter, if dark matter and the ICL are related, we should expect a correlation between the two.  
In Figure~\ref{Fig_ICL} we show as yellow contours the DM distribution from our Model-2 while in blue we show the contours for the DM distribution from our Model-3 (Model-1 is not shown but it is very similar to Model-2). The DM component is obtained after subtracting the mass associated to the galaxies from the total mass. In general we find good correspondence between the distribution of the ICL and the two DM models. \\

\begin{figure} 
   \includegraphics[width=9cm]{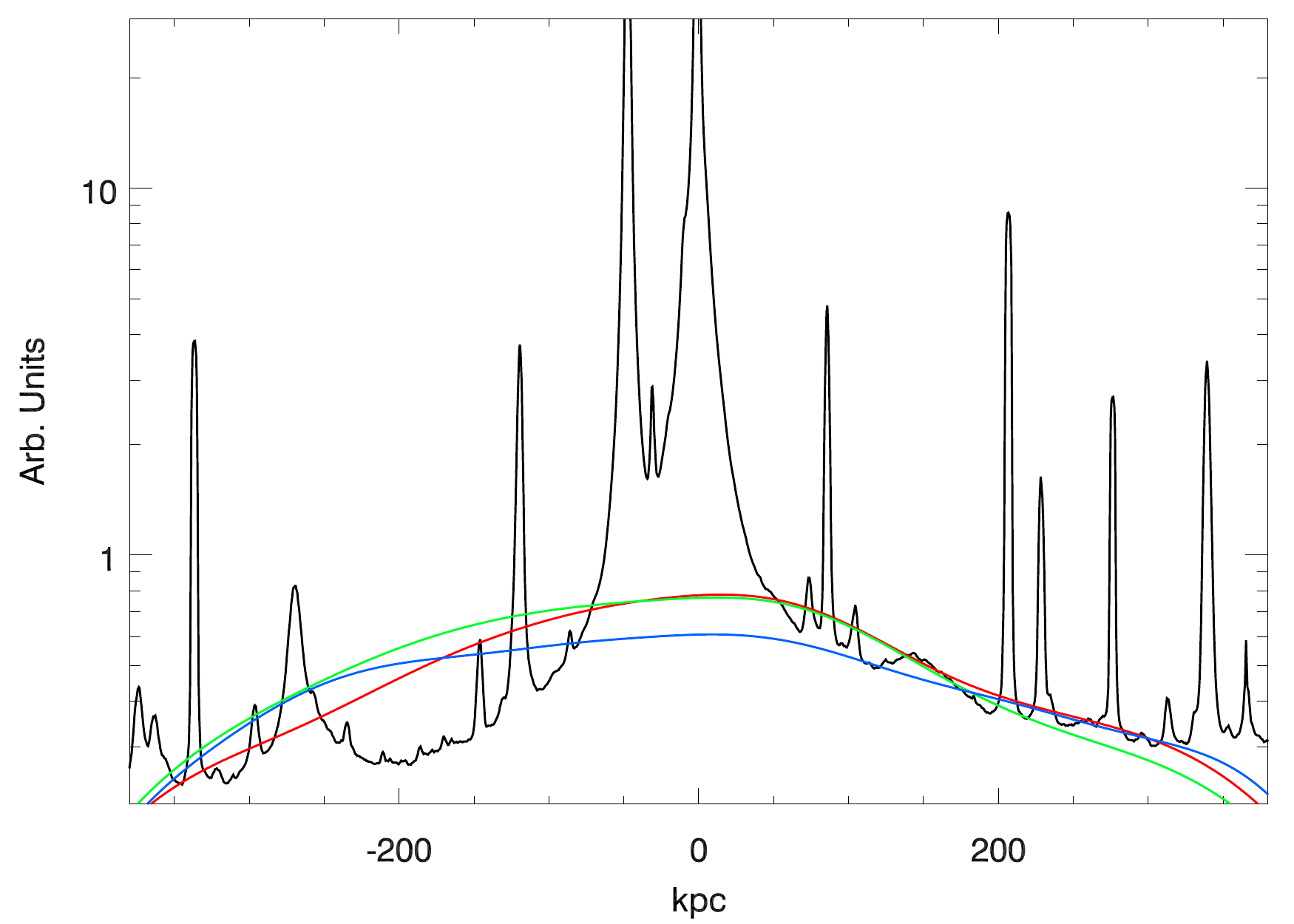}
      \caption{One dimensional scan of the light distribution vs DM. The x-axis is the distance to the BCG. The solid black line shows the mean of the light emission in the F277W band along the straight line shown in Figure~\ref{Fig_ICL}. The mean is computed over a box of 10x10 pixels at each position. The colored lines are the corresponding mean of the DM component for the three lens models discussed in this work. The red color is for the driver model, the green line is for Model-2, or low-resolution (20x20 grid points) with 25 systems, and the blue model is for the high-resolution (25x25 grid points) Model-3 with 25 systems. 
      The DM profiles are re-scaled by arbitrary units to visually match the profile of the light emission. See Figure~\ref{Fig_All} for a direct comparison of the profiles without the re-scaling. 
         }
         \label{Fig_ICL_scan}
\end{figure}

A more quantitative comparison is shown in Figure~\ref{Fig_ICL_scan} where we compute the average of the ICL or the DM along a straight line. This line is shown in Figure~\ref{Fig_ICL} and it intersects the ICL from west to east, passing through the central BCG in the middle. The average is computed at each position as the mean over a box of size $0.18"\times0.18"$ and centered in the line. The black curve in Figure~\ref{Fig_ICL_scan} corresponds to the light distribution. The colored lines are for the driver model or  Model-1 (red), Model-2 (green), and Model-3 (blue). The curves for the DM models have been re-scaled by an arbitrary number to match the black curve.

In the east part of the cluster we find good correspondence between all three models and the ICL. This is not true in the west part of the cluster, where the cavity clearly seen in the ICL at $\approx -200$ kpc in Figure~\ref{Fig_ICL_scan} is not observed in any of the DM models. 

\begin{figure} 
   \includegraphics[width=9cm]{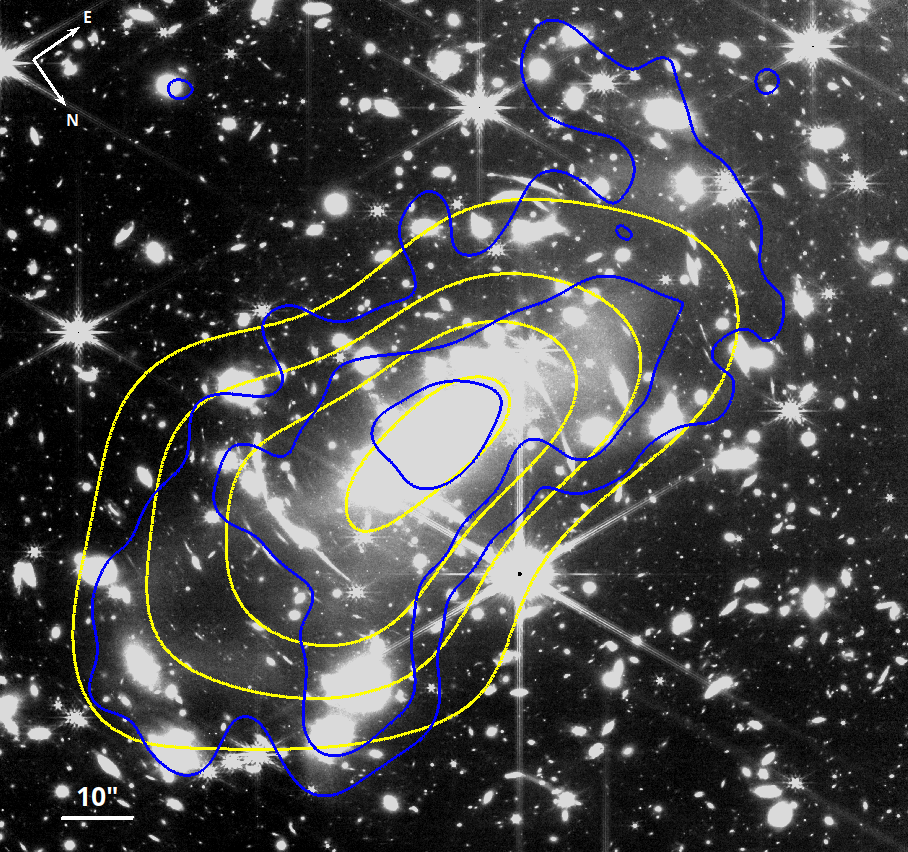}
      \caption{
      Comparison of the dark matter and globular cluster distribution (number density). The image corresponds to the F356W JWST filter. Yellow contours are the smooth component of the dark matter distribution (Model-2), while blue contours are for a Gaussian filtered version (FWHM=1.5") of the distribution of globular clusters from \cite{Lee2022}.   
         }
         \label{Fig_DMvsGC}
\end{figure}

In addition to the ICL, another possible tracer of the potential are globular clusters, whose distribution could correlate with the distribution of dark matter, since as the stars in the ICL, globular clusters respond to gravitational forces. The superior sensitivity and spatial resolution of JWST allows to detect these clusters with unprecedented detail. Preliminary results based on JWST data in SMACS0723 are presented in \cite{Lee2022,Faisst2022}. It is interesting to compare our results with those from earlier work. 
Figure \ref{Fig_DMvsGC} compares the observed ICL in the F356W filter with the distribution of dark matter (yellow contour) and the distribution of globular clusters (blue contours) from \cite{Lee2022}. To compute the blue contours we have smoothed the distribution of globular clusters with a Gaussian of FWHM=1.5". To first order, there is a good  spatial correspondence between the DM, ICL and globular cluster distribution, with all three components centered in the same point (BCG) and having similar alignments in the east-west direction.  As in the case of the ICL, the distribution of globular clusters appears to show a similar deficit in number density at the position of the cavity on the west side of the cluster. This cavity has no correspondence in the distribution of DM.

In terms of radial profiles, we show a comparison of our lens model with the ICL profiles from \cite{Montes2022} and the globular cluster profile from \cite{Lee2022} in Figure~\ref{Fig_All}. For the globular clusters, we have re-scaled the surface number density (expressed as number per kpc$^2$) by a factor $2\times 10^9$ in order for the resulting profile to overlap with the ICL profile. For comparison we plot a power law $R^{-1.3}$ as a dashed line. This power law reproduces well the profile of the ICL and the globular cluster number density. \\

The mass profiles from the three lens models are shown as a solid lines. Within the inner 20 kpc region, the total mass and the ICL have similar profiles. This is expected in our lens model since the compact component of the lens model takes directly the light distribution of member galaxies, including the central BCG. Since near the centre of the BCG, the bulk of the mass is expected to come from stars (or the baryonic component in general), by comparing our lens model with the ICL profile from \cite{Montes2022}, we find that either i) there is $\approx 10$ times more dark matter than stellar mass within the central 20 kpc or ii) the stellar mass from the ICL is underestimated by some factor.

Beyond $\approx 20$ kpc, the total mass profile is clearly shallower than the profile of the ICL and the number density of globular clusters. This departure is interesting and needs to be studied in other clusters with more constraints. Increasing the number of lensing constraints will allow to improve the spatial resolution of the lens model.


\section{Discussion and conclusions}\label{Sect_discussion}

The new data from JWST reveals a wealth of new candidate lensed  galaxies. Future observations of these candidates will secure their redshifts, which can then be compared with the geometric redshift estimate based on our driver model. If spectroscopic confirmation validates the method of estimating distances through geometric redshifts, future observations by the JWST can take advantage of a similar technique, where a handful of spectroscopic lensed galaxies may suffice to calibrate a lens model for distance estimation. Recent work has shown how photometric redshifts can predict erroneous redshifts for the case of dropout galaxies in the JWST bands \citep{Harikane2022,Naidu2022,Zavala2022}. An independent estimation of the distance to these galaxies can help reduce the uncertainty in the estimation of the redshift, and identify those galaxies that have large photometric redshifts (z>10) yet they are predicted by the lens model to be at much lower redshift.

Lens models like the one presented in this work are also needed to interpret sources near caustics. In the case of SMACS0723, \cite{Pascale2022} discuss a small pair of knots in the middle of the merging pair of images of system 5 (see their Figure 2). Since the lens model has a resolution comparable to the separation between the knots in the pair, the magnification in these knots is better estimated by interpolating the magnification.  Based on symmetry arguments, the critical curve must pass between these two points, so they are equidistant to it ($d\approx0.08")$. Since the magnification near a fold caustic scales as $\mu = A/d$ \citep{SchneiderBook1992}, we can estimate $A$ from our lens model. We find $A\approx 58"$, which results in $\mu\approx 725$ for each one of the images in the pair. This estimate matches very well the value quoted in \cite{Pascale2022} of $\mu\approx 750$.

The greater sensitivity of JWST to the ICL offers new opportunities to study the correlation between the DM and ICL. In addition, the improved spatial resolution in the infrared bands allows for detection of small clumps of old stellar populations in the cluster stripped from their hosts galaxies. The first image of JWST on this cluster reveals hundreds of unresolved clumps that are interpreted as globular clusters \citep{Lee2022,Faisst2022}, but could be also the surviving remnants (after a close encounter with a larger galaxy in the cluster, such as the BCG) of compact galactic cores.  Both the stars in the ICL and the globular clusters are expected to interact with the rest of the matter in the cluster mostly through gravitational forces, and hence behave similar to dark matter. 

\begin{figure} 
   \includegraphics[width=9cm]{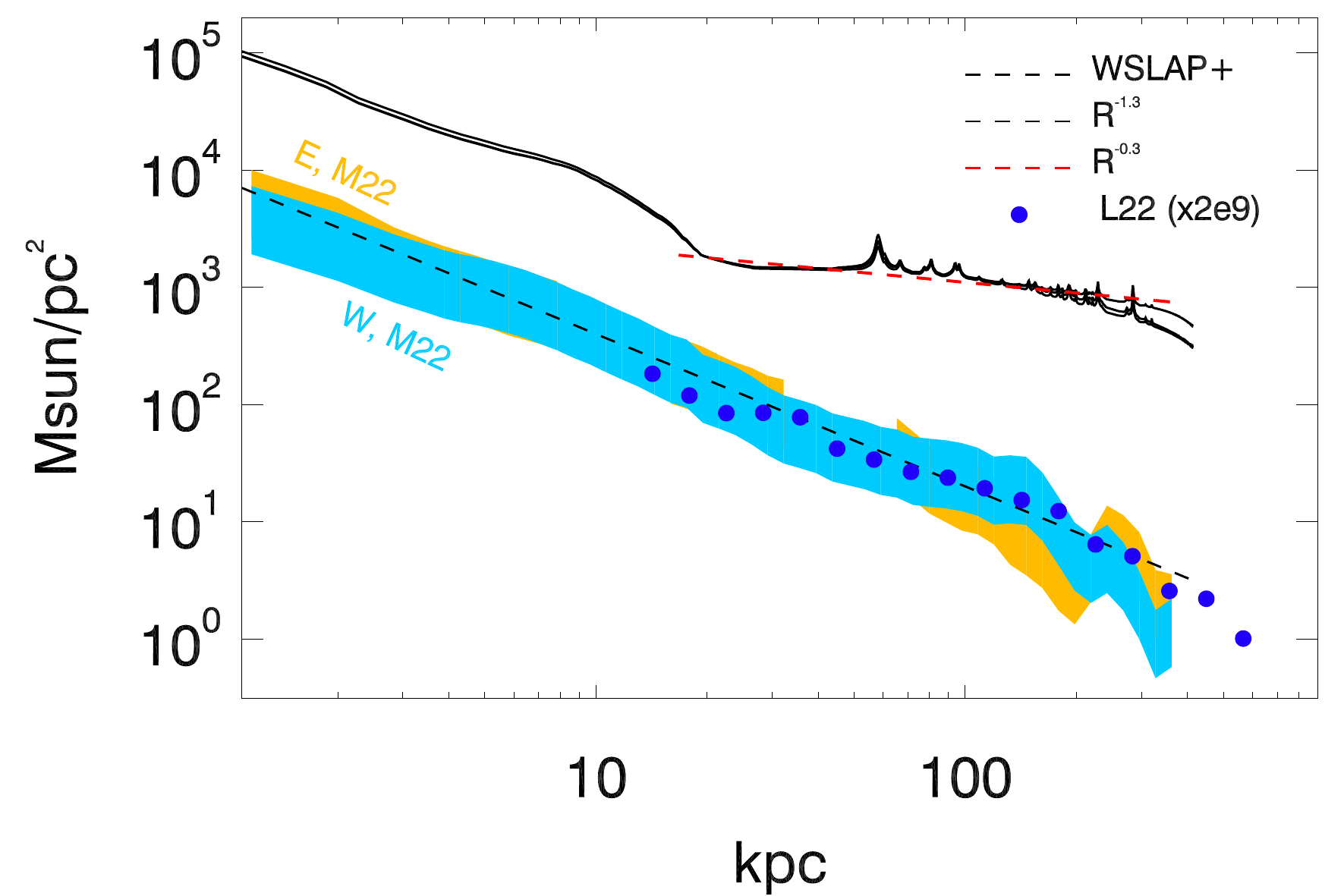}
      \caption{
      Comparison of the total mass profile from the three lens models (solid lines) with the ICL profile from \cite{Montes2022} (shaded orange and blue regions for the East and West sectors respectively) and the globular cluster number profile from \cite{Lee2022} (blue dots). For the later, we have re-scaled the number density by an arbitrary number of $2\times 10^9$ in order to overlap with the ICL profile. The black dashed line is a power law that scales with distance as $R^{-1.3}$. The red dashed line is a power law that scales as $R^{-0.3}$. 
         }
         \label{Fig_All}
\end{figure}

In this work we use a free-form modelling technique which makes minimal assumptions about the distribution of dark matter, and find that in general the DM traces well the ICL and globular cluster distribution but we also find that the small loop-like structures (and associated cavities) to the east and west of the central region of SMACS0723 have no obvious correspondence in the DM distribution. At distances from the centre comparable to the Einstein radius ($\sim 100$ kpc, and hence well constrained by the available data) we find that the dark matter profile is significantly shallower than the ICL and globular cluster distributions. This is also found in simulation of galaxy clusters. In \cite{Asensio2020}, the authors analyze the EAGLE simulations and find that the ICL profile is steeper than the total mass profile. In particular, they find that the ratio between the ICL and total mass profiles is a power law with slope $-1$. Interestingly, in the range between $\approx 20$ kpc and $\approx 200$ kpc we find a similar ratio between the total mass and ICL (and globular cluster) profiles, with the ICL and globular cluster profiles falling as $\sim R^{-1.3}$ while the total mass falls as $\sim R^{-0.3}$ (see Figure~\ref{Fig_All}). Similar conclusions are found in \cite{Pillepich2018} where, based on the IllustrisTNG simulations, the 3D profile of the ICL in massive clusters is found to fall faster (by approximately 1 dex from their Figure 6 at around 100 kpc distance) than the canonical NFW profile commonly used to describe dark matter profiles. Earlier work based on the EAGLE simulations shows a similar trend \citep{Schaller2015}. Hence, our results on SMACS0723 are in agreement with the ones derived from N-body simulations. 

We observe differences in the range $\approx 20$--$200$ kpc between the ICL (and globular cluster) profiles and the total mass (mostly dominated by dark matter in this distance range). We speculate that this may be related to the different formation times of the cluster dark matter halo and ICL. Since dark matter is more loosely bounded to their host halos (as it mostly resides on the outskirts of the galaxies, with the central region being more baryon dominated), it can be stripped more easily during the first encounters with the cluster and hence retaining the initial (relatively large) angular momentum. The baryonic component (stars in our case) is more concentrated around the centre of the satellite galaxies and can survive more encounters with the cluster, and without being stripped away. In each encounter, the satellite galaxy looses angular momentum due to dynamical friction and can get closer to the BCG \citep{Contini2018,Chun2022}. Stars that are stripped at a later time lose part of their bulk kinetic energy this way, and when stripped from their hosts can remain at shorter radii, resulting in profiles that are steeper (more concentrated) than the dark matter profiles. Globular clusters and galactic core remnants are subject also to dynamical friction and hence expected to orbit closer to the BCG, resulting in more concentrated profiles. N-body simulations also show how the radial distribution of subhalos is steeper than the distribution of dark matter \citep{Gao2004}.

The presence of cavities in the distribution of the ICL, and not detected in the total mass distribution, is another interesting difference. The formation of cavities in the ICL but not in the DM distribution could be due to the different distribution of stars and dark matter inside the satellite galaxies before they enter the galaxy cluster, and the striping mechanism starts to take place. The dark matter, forming an extended halo around the satellite galaxy, is easily tidally stripped from its host galaxy as it enters the cluster and starts orbiting around the BCG.  The better ability of the baryonic matter to cool down more efficiently and form more concentrated structures like disks or bulges  facilitates the survival of the bulge (or disk)  as they orbit the minimum of the potential. During a close encounter with the BCG, parts of the bulge or disc of a satellite galaxy can be tidally stripped, creating the loop-like structures  and associated cavities. Tidal stripping of satellite galaxies has been claimed as responsible for filamentary structures seen in the ICL of the nearby Virgo cluster \citep{Mihos2005}. Structures that resemble the loop-cavity system are also observed in nearby galaxies that had recent encounters with satellite galaxies  \citep{MartinezDelgado2022}. In simulations, faint structures in the ICL, that resemble the loop-cavity structures can be appreciated in Figures 3 and 4 in \cite{Pillepich2018}.

Perhaps one of the most interesting findings is the connection between the ICL and globular cluster distribution, both having a  similar profile. This connection could be easily explained if the ICL corresponds to the outer envelopes of the alleged globular clusters. In this case, the globular clusters should be re-interpreted as the surviving galactic cores of the infalling satellite galaxies. 

More examples like the one studied in this work are needed in order to extract a firmer conclusion regarding the connection between the ICL, globular cluster and DM distributions. In particular, the addition of new constraints (with confirmed spectroscopic redshift) will allow us to increase the resolution of the lens model, revealing perhaps finer details in the distribution of DM that can not be unveiled with the current set of constraints. For the particular case of SMACS0723, the number of lensing constraints around the west cavity is very small ($\approx 4$ lensed galaxies in this region). Future analysis based on JWST data, especially of low redshift clusters for which both ICL and globular clusters are more easily detected, and with abundant lensing constraints (such as the Hubble Frontier Fields Clusters) will enable more precise conclusions on the correlation between the ICL, globular cluster and DM distributions.

\begin{acknowledgements}
 J.M.D. acknowledges the support of project PGC2018-101814-B-100 (MCIU/AEI/MINECO/FEDER, UE) Ministerio de Ciencia, Investigaci\'on y Universidades.  This project was funded by the Agencia Estatal de Investigaci\'on, Unidad de Excelencia Mar\'ia de Maeztu, ref. MDM-2017-0765. 
 M.P. was funded through the NSF Graduate Research Fellowship grant No. DGE 1752814.
 A.Z. acknowledges support by Grant No. 2020750 from the United States-Israel Binational Science Foundation (BSF) and Grant No. 2109066 from the United States National Science Foundation (NSF), and by the Ministry of Science \& Technology, Israel.
 G.M. acknowledges funding from the European Union’s Horizon 2020 research and innovation programme under the Marie Sklodowska-Curie grant agreement No MARACHAS-DLV-896778.
 M.J. is supported by the United Kingdom Research and Innovation (UKRI) Future Leaders Fellowship `Using Cosmic Beasts to uncover the Nature of Dark Matter' (grant number MR/S017216/1). 
 M.G.L.  was supported by the National Research Foundation grant funded by the Korean Government (NRF-2019R1A2C2084019).
 M.M. acknowledges the Project PCI2021-122072-2B, financed by MICIN/AEI/10.13039/501100011033, and the European Union “NextGenerationEU”/RTRP.

\end{acknowledgements}

\bibliographystyle{aa} 
\bibliography{MyBiblio} 

\newpage 

\begin{appendix}

\section{Arc positions and redshifts}
This appendix presents all arc system candidates used in this work. The table is a compilation of systems presented in \cite{Mahler2022,Pascale2022,Caminha2022}. The last two rows are the positions of the two critical points used as extra constraints.

IDs of systems is shown in column one. Columns two and three give RA, DEC positions in degrees.  Column four lists the spectroscopic redshifts when available. Spectroscopic redshifts are marked in bold face. Column five lists the redshifts predicted by the driver model. In this case, errors correspond to the 68\% interval of the PDF. For all systems, only the first arc is given with its redshift. Colums six, seven, and eight give the original ID in Pascale et al. (2022), Caminha et al. (2022), and Mahler et al. (2022) respectively. \\
\ddag While finishing this paper, \cite{Noirot2022} published spectroscopic redshifts of some galaxies in the field of SMACS0723 including a redshift for our system 4, for which they find $z_{spec}=2.211$, in good agreement ($2\sigma$) with our geometric redshift estimate ($z_{geo}=2\pm0.1$). No other redshifts are published for the remaining arcs without spectroscopic redshifts.

%

\onecolumn
\begin{longtable}{|l|ll|ll|lll|}
    \caption{Arc positions and redshifts.} \label{tab_Arcs} \\
\hline \multicolumn{1}{|c|}{\textbf{ID}} & 
\multicolumn{1}{c}{\textbf{RA}} & 
\multicolumn{1}{c|}{\textbf{DEC}} &
\multicolumn{1}{c}{\textbf{z$_{\rm s}$}} &
\multicolumn{1}{c|}{\textbf{z$_{\rm m}$}} &
\multicolumn{1}{c}{\textbf{IDP}} &
\multicolumn{1}{c}{\textbf{IDC}} &
\multicolumn{1}{c|}{\textbf{IDM}} \\ \hline 
\endfirsthead

\multicolumn{7}{c}%
{{\bfseries \tablename\ \thetable{} -- continued from previous page}} \\

\hline \multicolumn{1}{|c|}{\textbf{ID}} & 
\multicolumn{1}{c}{\textbf{RA}} & 
\multicolumn{1}{c|}{\textbf{DEC}} &
\multicolumn{1}{c}{\textbf{z$_{\rm s}$}} &
\multicolumn{1}{c|}{\textbf{z$_{\rm m}$}} &
\multicolumn{1}{c}{\textbf{IDP}} &
\multicolumn{1}{c}{\textbf{IDC}} &
\multicolumn{1}{c|}{\textbf{IDM}} \\ \hline 

\endhead

\hline 
\endfoot

\hline \hline
\endlastfoot

1.1  & 110.8407240  &  -73.4510787  &  {\bf 1.449}   &  1.45$\pm0.07$ & 1.1  &  2a  &  1.1  \\
1.2  & 110.8429489  &  -73.4548399  &    --   &  --   &  1.2  & 2b  &  1.2  \\
1.3  & 110.8389887  &  -73.4587844  &    --   &  --  &  1.3  & 2c  &  1.3  \\
\hline
2.1  & 110.8387288  &  -73.4510508  &  {\bf 1.3779}  & 1.37$\pm0.06$ &   2.1  &  3a  &  2.1  \\
2.2  & 110.8407771  &  -73.4552122  &    --    &  --  &  2.2  &  3b  &  2.2  \\
2.3  & 110.8364983  &  -73.4588136  &    --    &  --  &  2.3  &  3c  &  2.3  \\
\hline
3.1  & 110.8304431  &  -73.4485622  &  {\bf 1.9914}  &  1.97$\pm0.11$   & 3.1  &  1a  &  3.3  \\
3.2  & 110.8318194  &  -73.4552311  &    --   &  --  &  3.2  &  1b  &  3.2  \\
3.3  & 110.8252159  &  -73.4596604  &    --   &  --  &  3.3  &  1d  &  3.1  \\
3.4  & 110.8232656  &  -73.4548634  &    --   &  --  &  3.4  &  1c  &  3.4  \\
\hline
4.1  & 110.8069982  &  -73.4584308  &  \ddag   &  2.00$\pm0.10$   &  4.2  &  4c  &  4.2  \\
4.2  & 110.8052367  &  -73.4546325  &   --   &  --   &  4.1  &  4b  &  4.1  \\
4.3  & 110.8132881  &  -73.4487869  &   --   &  --   &  4.3  &  4a  &  4.3  \\
\hline
5.1  & 110.8238908  &  -73.4518820  &  {\bf 1.425}   &  1.45$\pm0.06$  & 5.1  &  6a  &  5.1  \\
5.2  & 110.8223529  &  -73.4527831  &   --   &  --   &  5.2  &  6b  &  5.2  \\
5.3  & 110.8209254  &  -73.4602058  &   --   &  --   &  5.3  &  6c  &  5.3  \\
\hline
6.1  & 110.8358540  &  -73.4518199  &   --   &  1.67$\pm0.07$   &  6.1  &  9a  &  7.1  \\
6.2  & 110.8367611  &  -73.4530868  &   --   &  --   &  6.2  &  9b  &  7.2  \\
6.3  & 110.8303933  &  -73.4608436  &   --   &  --   &       &  9c  &  7.3  \\
\hline
7.1  & 110.7947604  &  -73.4490975  &  {\bf 5.1727}    &  5.21$^{+5.1}_{-0.6}$  & 7.2  &      &  17.1  \\
7.2  & 110.7954442  &  -73.4487211  &   --   &  --   &  7.1  &      &  17.2  \\
7.3  & 110.7996039  &  -73.4470866  &   --   &  --   &  7.3  &      &  17.3  \\
\hline
8.1  & 110.8023784  &  -73.4602055  &   --    & 8.06$^{+1.7}_{-0.8}$   &  8.1  &  7a  &  12.1  \\
8.2  & 110.7995598  &  -73.4553501  &   --    &  --  &  8.2  &  7b  &  12.2  \\
8.3  & 110.8130564  &  -73.4466651  &   --    &  --  &  8.3  &  7c  &     \\
\hline
9.1  & 110.8050637  &  -73.4589656  &   --    &  2.62$\pm0.14$    &  9.2  &  10c  &  13.3  \\
9.2  & 110.8028896  &  -73.4549564  &   --    &  --  &  9.1  &  10b  &  13.2  \\
9.3  & 110.8127004  &  -73.4481250  &   --    &  --  &  9.3  &       &  13.1  \\
\hline
10.1 & 110.8235289  &  -73.4517392  &   --    &  1.45$\pm0.06$    &  10.1  &      &    \\
10.2 & 110.8216192  &  -73.4528243  &   --    &  --  &  10.2  &      &    \\
10.3 & 110.8205119  &  -73.4601152  &   --    &  --  &  10.3  &      &    \\
\hline
11.1 & 110.8107306  &  -73.4569574  &   --    &   1.47$^{+8.1}_{-0.2}$   &  11.2  &      &  19.1  \\
11.2 & 110.8101464  &  -73.4561599  &   --    &  --  &  11.1  &      &  19.2  \\
\hline  
12.1 & 110.8221364  &  -73.4491504  &   --    &   1.66$\pm0.06$   &  12.1  &  13a  &  14.1  \\
12.2 & 110.8146179  &  -73.4544119  &   --    &  --  &  12.2  &  13b  &  14.2  \\
12.3 & 110.8173093  &  -73.4593170  &   --    &  --  &  12.3  &  13c  &  14.3  \\
\hline
13.1 & 110.8297224  &  -73.4489907  &   --    &  3.02$^{+0.26}_{-0.17}$  &  13.1  &  12a  &  6.3  \\
13.2 & 110.8219150  &  -73.4542067  &   --    &  --  &  13.2  &  12c  &  6.2  \\
13.3 & 110.8231150  &  -73.4617081  &   --    &  --  &  13.3  &  12d  &  6.4  \\
13.4 & 110.8324286  &  -73.4544642  &   --    &  --  &  13.4  &  12b  &  6.1  \\
\hline
14.1 & 110.8015568  &  -73.4583546  &   --    &  --  &  14.1  &       &    \\
14.2 & 110.8018148  &  -73.4589480  &   --    &  --  &  14.2  &       &    \\
14.3 & 110.8022270  &  -73.4590843  &   --    &  --  &  14.3  &       &    \\
\hline
15.1 & 110.8193895  &  -73.4487436  &   --    &  1.82$\pm0.08$   &  15.1  &  11a  &    \\
15.2 & 110.8113813  &  -73.4546235  &   --    &  --  &  15.2  &  11b  &    \\
15.3 & 110.8139705  &  -73.4590522  &   --    &  --  &  15.3  &  11c  &    \\
\hline
16.1 & 110.8206200  &  -73.4527181  &   --    &  1.26$\pm0.05$  &  16.1  &       &  11.1  \\
16.2 & 110.8205250  &  -73.4528156  &   --    &  --  &  16.2  &       &  11.2  \\
16.3 & 110.8207626  &  -73.4597746  &   --    &  --  &        &       &  11.3  \\
\hline
17.1 & 110.8239479  &  -73.4575528  &   --    &  2.33$\pm0.11$  &  18.2  &  8c   &  8.2  \\
17.2 & 110.8231354  &  -73.4558083  &   --    &  --  &  18.1  &  8b   &  8.1  \\
17.3 & 110.8297769  &  -73.4474619  &   --    &  --  &  18.3  &  8a   &  8.3  \\
\hline
18.1 & 110.8216711  &  -73.4506362  &   --    &  1.36$\pm0.05$  &  19.1  &       &  9.1  \\
18.2 & 110.8167450  &  -73.4537968  &   --    &  --  &  19.2  &       &  9.2  \\
18.3 & 110.8179340  &  -73.4590101  &   --    &  --  &  19.3  &       &  9.3  \\
\hline
19.1 & 110.8208804  &  -73.4507461  &  {\bf 1.3825}  &  1.37$\pm0.05$   &    &  5a       &  10.1  \\
19.2 & 110.8164058  &  -73.4535733  &   --    &  --  &    &  5b       &  10.2  \\
19.3 & 110.8173046  &  -73.4589942  &   --    &  --  &    &  54       &  10.3  \\
\hline
20.1 & 110.8165814  &  -73.4519445  &   --    &   1.20$^{+3.4}_{-0.07}$  &    &  14a      &    \\
20.2 & 110.8159392  &  -73.4523932  &   --    &  --  &    &  14b      &    \\
\hline  
21.1 & 110.8168354  &  -73.4485770  &   --    &   2.19$\pm0.11$   &    &  15a      &    \\
21.2 & 110.8086654  &  -73.4541442  &   --    &  --  &    &  15b      &    \\
21.3 & 110.8115827  &  -73.4596446  &   --    &  --  &    &  15c      &    \\
\hline
22.1 & 110.8293400  &  -73.4561204  &   --    &   2.27$^{+0.73}_{-0.22}$  &    &  16a      &    \\
22.2 & 110.8268630  &  -73.4578161  &   --    &  --  &    &  16b      &    \\
\hline  
23.1 & 110.8258363  &  -73.4502839  &   --    &   1.59$\pm0.06$   &    &           &  15.1  \\
23.2 & 110.8201612  &  -73.4539789  &   --    &  --  &    &           &  15.2  \\
23.3 & 110.8213975  &  -73.4602314  &   --    &  --  &    &           &  15.3  \\
\hline
24.1 & 110.8085708  &  -73.4494083  &   --    &  2.10$\pm0.12$   &    &           &  16.1  \\
24.2 & 110.8019579  &  -73.4526322  &   --    &  --  &    &           &  16.2  \\
24.3 & 110.8058921  &  -73.4595997  &   --    &  --  &    &           &  16.3  \\
\hline
25.1 & 110.7927038  &  -73.4484814  &   --    &   2.16$^{+0.87}_{-0.23}$   &    &           &  18.1  \\
25.2 & 110.7936842  &  -73.4482439  &   --    &  --  &    &           &  18.2  \\
25.2 & 110.7964129  &  -73.4469406  &   --    &  --  &    &           &  18.3  \\
\hline
26.1 & 110.7917089  &  -73.4566332  &   --    &  --  &    &           &  20.1  \\
26.2 & 110.7914913  &  -73.4558973  &   --    &  --  &    &           &  20.2  \\
\hline
27.1 & 110.8032246  &  -73.4582886  &   --    &   2.81$^{+0.22}_{-0.15}$   &    &           &  21.1  \\
27.2 & 110.8041292  &  -73.4531883  &   --    &  --  &    &           &  21.2  \\
27.3 & 110.8136692  &  -73.4495378  &   --    &  --  &    &           &  21.3  \\
\hline
CP5 & 110.8234036   &  -73.4522538  &   {\bf 1.425}    &  & &         & \\
CP7 & 110.7951736   &  -73.4488732  &   {\bf 5.1727}   &  & &         & \\
\end{longtable}
\twocolumn


\end{appendix}

\end{document}